# SIMULATION OF SPECIAL BUBBLE DETECTORS FOR PICASSO

G. Azuelos†, M. Barnabé-Heider†, E. Behnke§, K. Clark‡, M. Di Marco†, P. Doane†, W. Feighery§, M-H. Genest†, R. Gornea†, R. Guenette†, S. Kanagalingam*, C. Krauss‡, C. Leroy†, L. Lessard†, I. Levine§, J. P. Martin†, A. J. Noble‡, R. Noulty*, S. N. Shore¶, U. Wichoski†, V. Zacek†

†Groupe de Physique des Particules, Département de Physique, Université de Montréal, C.P. 6128, Succ.Centre-Ville, Montréal (Québec) H3C 3J7, Canada

‡Department of Physics, Queens University

Kingston (Ontario) K7L 3NG, Canada

§Department of Physics and Astronomy, Indiana University South Bend

South Bend, Indiana, 46634, USA

¶Dipartimento di Fisica "Enrico Fermi",Universtità di Pisa,Pisa,I-56127,Italia

*Bubbles Technology Industries, Chalk River (Ontario) K0J 1J0, Canada

The PICASSO project is a cold dark matter (CDM) search experiment relying on the superheated droplet technique. The detectors use superheated freon liquid droplets (active material) dispersed and trapped in a polymerized gel. This detection technique is based on the phase transition of superheated droplets at about room temperature and ambient pressure. The phase transition is induced by nuclear recoils when an atomic nucleus in the droplets interacts with incoming subatomic particles. This includes CDM particles candidate as the neutralino (a yet-to-discover particle predicted in extensions of the Standard Model of particle physics). Simulations performed to understand the detector response to neutrons and alpha particles are presented along with corresponding data obtained at the Montreal Laboratory.

INTRODUCTION

Superheated droplet detectors, here referred to as special bubble detectors (SBD's), use superheated liquid droplets (active medium) dispersed and suspended in a polymerized gel[1,2]. Presently, these droplet detectors consist of an emulsion of droplets of a superheated liquid (such as $C_3F_8$, $C_4F_{10}$, $CF_3Br$, $CCl_2F_2$) which are metastable at about room temperature and ambient pressure. During the fabrication process, the droplets are dispersed in an aqueous solution to which a heavy salt (e.g. CsCl) is added to equalize densities so that the droplets neither sink nor raise to the surface. As a result of the fabrication process, a gel matrix with an isotropic distribution of droplets is obtained. The operation of the detector can be controlled by applying an adequate pressure which makes the boiling temperature of the liquid raise, preventing bubble formation. Under this external pressure, the detectors are insensitive to radiation.

By removing the external pressure, the liquid becomes superheated and sensitive to incoming particles and radiation. Bubble formation occurs through liquid-to-vapour phase transitions, triggered by the energy deposited by incoming particles and radiation. The energy deposition can be the result of nuclear recoils (through interactions with neutrons or other particles), or gamma and alpha particles. Bubble detectors are threshold detectors as a minimal energy deposition is needed to induce a phase transition. Their sensitivity to various types of radiation strongly depends on the operating temperature and pressure. The liquid-to-vapour transition is explosive in nature and is accompanied by an acoustic shock wave which can be detected with piezoelectric sensors[3]. The lifespan of the detectors is very long (> 2 years) as they are re-usable by re-compressing the bubbles back to droplets.

BUBBLE FORMATION IN SBD'S

The response of the SBD's to incoming particles or radiation is determined by thermodynamic properties of the active liquid, such as operating temperature and pressure. The detector operation can be understood in the framework of the theory of Seitz[4] which describes bubble formation as being triggered by the heat spike produced by the energy deposited when a particle traverses a depth of superheated medium. Bubble formation will occur when a minimum deposited energy, $E_{Rth}$, exceeds the threshold value

$$E_c = \frac{16\pi}{3} \frac{\sigma(T)^3}{(p_i - p_0)^2}, \qquad (1)$$

within a distance $l_c \approx aR_c$, where the critical radius $R_c$ is given by:

$$R_c = \frac{2\sigma(T)}{(p_i - p_0)}, \qquad (2)$$

where $p_0$ and $p_i$, the externally applied pressure and the vapour pressure in the bubble, respectively, are temperature dependent. The surface tension of the liquid-vapour interface at temperature $T$ is given by $\sigma(T)=\sigma_0(T_c-T)/(T_c-T_0)$ where $T_c$ is the critical temperature of the gas, $\sigma_0$ is the surface tension at a reference temperature $T_0$, usually the boiling temperature $T_b$. The probability, $P(E_{dep}, E_{Rth})$, that a recoil nucleus at an energy near threshold will generate an explosive droplet-bubble transition is given by

$$P = 1 - \exp\left[\frac{-b(E_{dep} - E_{Rth})}{E_{Rth}}\right], \qquad (3)$$

where $b$ is a free parameter to be fitted to the data.

SIMULATION OF THE ALPHA RESPONSE

**Experimental alpha response**

The heavy salt and other ingredients, mixed in the gel at the present stage of detector production, contain contaminants which are $\alpha$-emitters, such as U/Th and their daughter nuclei. The $\alpha$ background is one of the main backgrounds at normal temperatures of SBD operation since other backgrounds contribute to the detector signal predominantly at higher temperatures. For instance, neutrons, discussed in the next section, should be an important background at operating temperatures. However, neutrons are effectively shielded against in the case of the PICASSO experiment by operating the detectors in a water shielding, deep underground[5]. Therefore, the response of SBD to α-particles has to be investigated carefully. The alpha response was studied using a 1$l$ SBD-1000 fabricated at BTI (Bubble Technology Industry). In the fabrication process, 27.8 m$l$ of an americium solution (AmCl$_3$ in 0,5M HCl) of known activity (0,7215 Bq/m$l$) were added to the gel solution to create a detector spiked with 20 Bqs of $^{241}$Am. Then, the detector liquid was added, followed by the mixing and polymerisation procedures. Using this detector, the alpha detection efficiency was measured at temperatures ranging from 6°C to 50°C (see Figure 1). The efficiency is temperature dependent. In the temperature range studied, the detector is essentially insensitive to $\gamma$ and $\beta$ radiation. Since the experimental data, shown in Figure 1, were obtained without any neutron shielding, the neutron background possibly has a non-negligible contribution to the alpha response at temperatures below ≈20°C. Another alpha response measurement is currently in progress using a neutron-shielded uranium-spiked detector. The preliminary results indicate that the alpha threshold may be higher than shown in Figure 1.

**Alpha response Monte Carlo**

The alpha response was simulated, using Geant 4.5.2[6]. The density of the superheated droplets was varied as a function of temperature and the droplets were dispersed randomly in the gel. For all simulations, the experimentally known distribution of droplet diameters was used. The loading was assumed to be 0.7%, a choice validated by a comparison between the neutron efficiency of the simulated detector and another 1$l$ detector for which the droplet distribution and loading were directly measured. Alpha particles were generated randomly in the gel, with an energy spectrum corresponding to the $^{241}$Am decay. The ionization of low energy nuclei was taken into account, using the ICRU_R49[7] nuclear stopping model and the SRIM2000p[8] electronic stopping model.

The minimal energy deposition needed to trigger a vaporization is known from the experimental threshold curves obtained with neutrons, assuming a head-on collision between a neutron and a nucleus inside the droplet. To maintain the consistency between neutron and alpha measurements, the simulations strongly suggest that all the recoiling nucleus energy is needed. The critical length for the necessary energy to be deposited will hence be greater than the range of the nucleus considered.

Monte Carlo studies of the alpha response indicate that the experimental efficiency is too high for the vaporization to be caused only by elastic collisions between alpha particles and nuclei in the droplets: the phase transition must hence be triggered by the alpha particles' ionization loss in the droplets. Since the americium solution used in the spiked detector fabrication is hydrophilic and since the freon droplets are hydrophobic, we can assume that the americium doesn't diffuse in the droplets. Furthermore, the experimental efficiency is low enough to consider no surfactant effect[9]. Under those assumptions, the contribution of the recoiling short-range nucleus can be neglected ($^{237}$Np in the case of $^{241}$Am decays).

Under the assumption that the recoiling nucleus triggering vaporization at neutron threshold is fluorine, the dE/dx required to trigger a phase transition is too high to explain the efficiencies seen in the alpha case. This is not completely understood. It suggests that the minimal energy deposited at neutron threshold must be defined by the carbon recoil.

Taking the probability function (Equation 3) and this minimal energy requirement, the critical length as a function of temperature and the value of *b* were deduced from the fit to the data (see Figure 1). The value of the critical length obtained is L=18$R_c$ and *b*=1. A 68.3% C.L interval for the Poisson signal was calculated for a simulated number of counts lower than 21. The error becomes √N for a larger number of events. It is possible that the discrepancy between the Monte Carlo and the experimental data below 20°C is due to a non-negligible neutron contribution, as mentioned earlier. The hypothesis is often made that a critical-radius spherical cavity must be reached to achieve bubble nucleation. However, the simulations results show that the energy must be deposited over effective track lengths greater than a critical diameter. Therefore, one can assume that a vapour cavity may initially extend along the particle track before quickly acquiring a spherical shape.

Due to the short range of alpha particles in the detector, the alpha response strongly depends on the size of the droplets dispersed in the gel. Simulations of a 1% loaded $^{241}$Am spiked detector were performed for three different droplet sizes (see Figure 2). The maximal alpha detection efficiency is shown to be inversely proportional to the droplet radius.

SIMULATION OF THE NEUTRON RESPONSE

**Experimental neutron response**

The SBD response to mono-energetic neutrons has been measured as a function of temperature. These mono-energetic neutrons are obtained from $^7Li(p, n)^7Be$ reactions at the Tandem van der Graaff facility of the Université de Montréal. The detector response (count rate) to mono-energetic neutrons of 200 and 400 keV as a function of operating temperature are shown in Figures 3 and 4, respectively, for a detector of 8 m*l* volume loaded with 100% $C_4F_{10}$ droplets (SBD-1000 detector).

**Neutron response Monte Carlo**

The response of an 8 m$l$ SBD-1000 detector to mono-energetic neutron beams was also simulated, using the same energy deposition and critical length requirements as in the alpha case. The loading was set to be 0.7% and, after analysis, the response was normalized by a multiplicative factor to fit the experimental data. This factor was interpreted as the loading correction from the initially assumed 0.7% value. The energy variation in the proton beam used to create neutrons causes the neutron energy to fluctuate around the beam energy mean value. Consequently, the neutrons generated in the simulation were given an initial energy with Gaussian fluctuations, e.g. σ=5 keV for 200 keV neutrons. Neutron elastic and inelastic scattering were considered for all types of nuclei using the ENDF/B-VI data library[10], in all detector parts. As can be seen in Figures 3 and 4, the simulated response fits well the experimental data for E = 200 and 400 keV, validating the critical length and minimal energy requirements used to fit the alpha response. Fitting the neutron experimental data also allows the determination of the detector's loading. The values of 0.62±0.04% and 0.68±0.02% obtained at 200 and 400 keV, respectively, are consistent with each other.

CONCLUSION

Simulations performed to understand the response of superheated droplet detectors to neutrons and alpha particles indicate that all the data can be well described with a unique, consistent set of variables which parameterize the underlying model of recoil energy threshold and energy deposition (Seitz theory). The simulations of the neutron data are able to trace the response of mono-energetic neutrons as a function of temperature over several orders of magnitude in count rate. The experimental alpha response is equally well described except at threshold, a discrepancy that could possibly be explained by a neutron contamination of the experimental data. Another alpha response measurement is under way using a neutron shielding. The simulations show that it is the alpha particles which mainly trigger droplet formation by their specific energy loss along the track, rather than the recoiling $^{237}$Np nucleus. The detection efficiency at higher temperatures is consistent with the assumption that the alpha emitters are distributed uniformly in the gel surrounding the droplets. Furthermore, the simulations show that the maximal alpha detection efficiency is inversely proportional to the droplet radius, indicating that a detector having higher droplet sizes would have a lower internal alpha background.

*Document*, Report BNL-NCS-17541 (ENDF-201), edited by P.F. Rose, National Nuclear Data Center, Brookhave National Laboratory, Upton, NY, USA (1991).

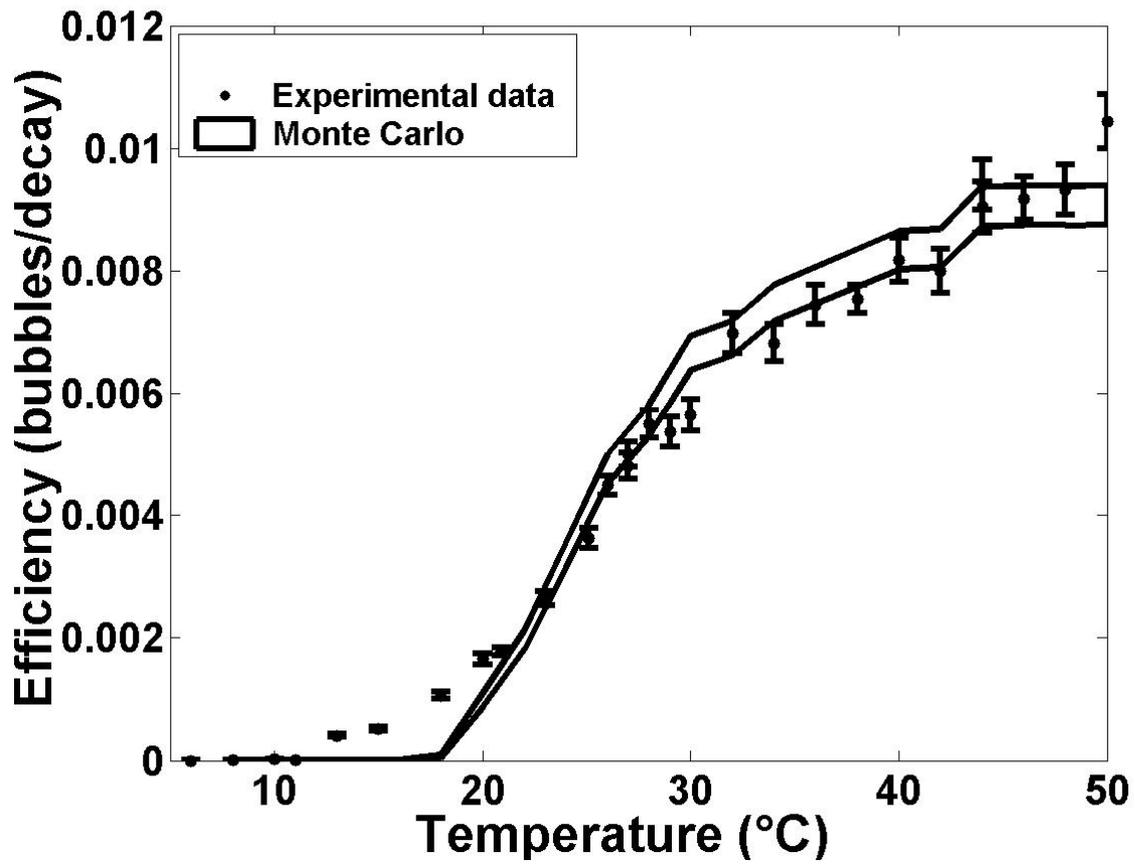

Figure 1: Detector response (count rate) of the 1*l* SBD-1000 spiked with 20 Bqs of $^{241}$Am as a function of temperature. A critical length of L=18$R_c$ is necessary to fit the data.

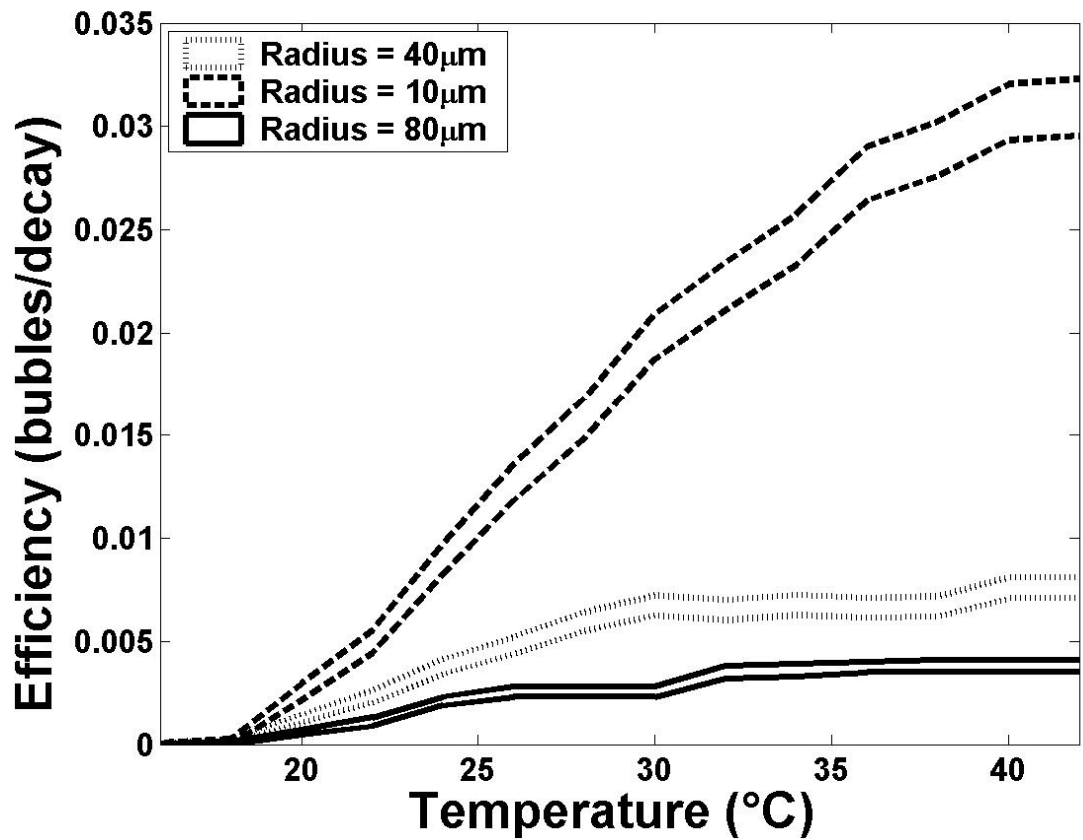

Figure 2: Simulation of a 1% loaded SBD-1000 detector spiked with 20 Bqs of $^{241}$Am for three different droplet sizes. The maximal alpha detection efficiency is inversely proportional to the droplet size.

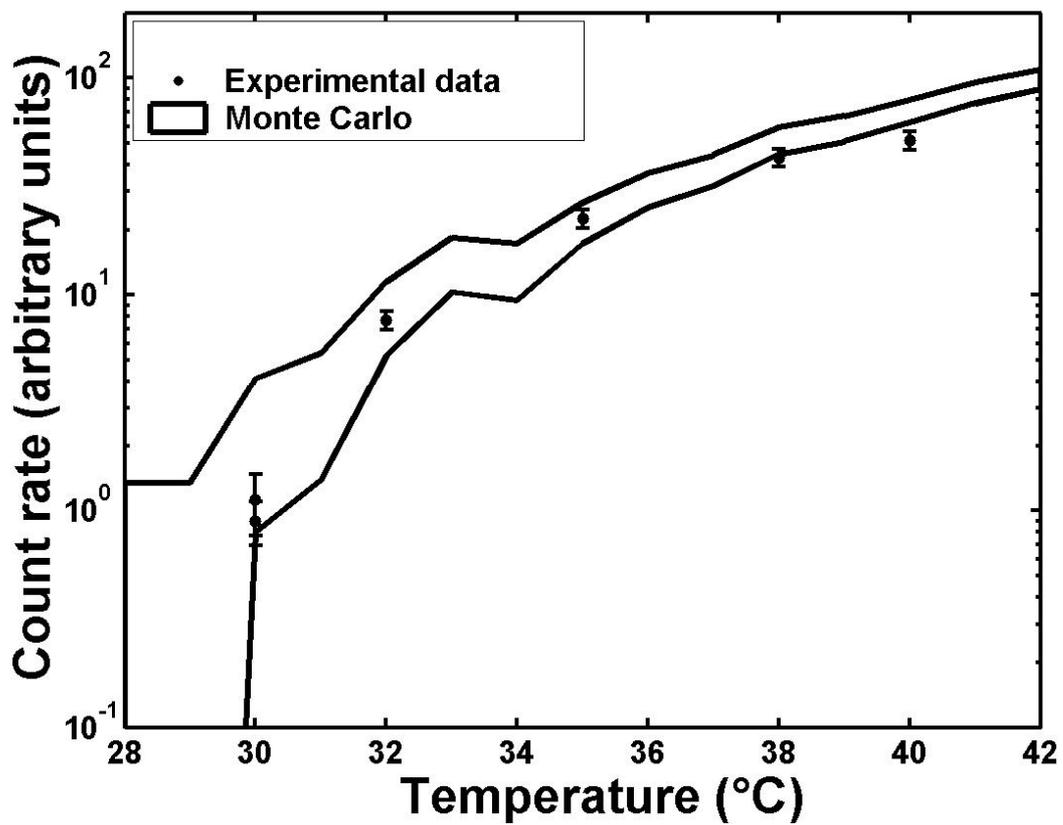

Figure 3: SBD-1000 response (count rate) to 200 keV neutrons as a function of temperature. The volume of the detector is 8 m*l*. The simulated response gives a loading of 0.62±0.04%.

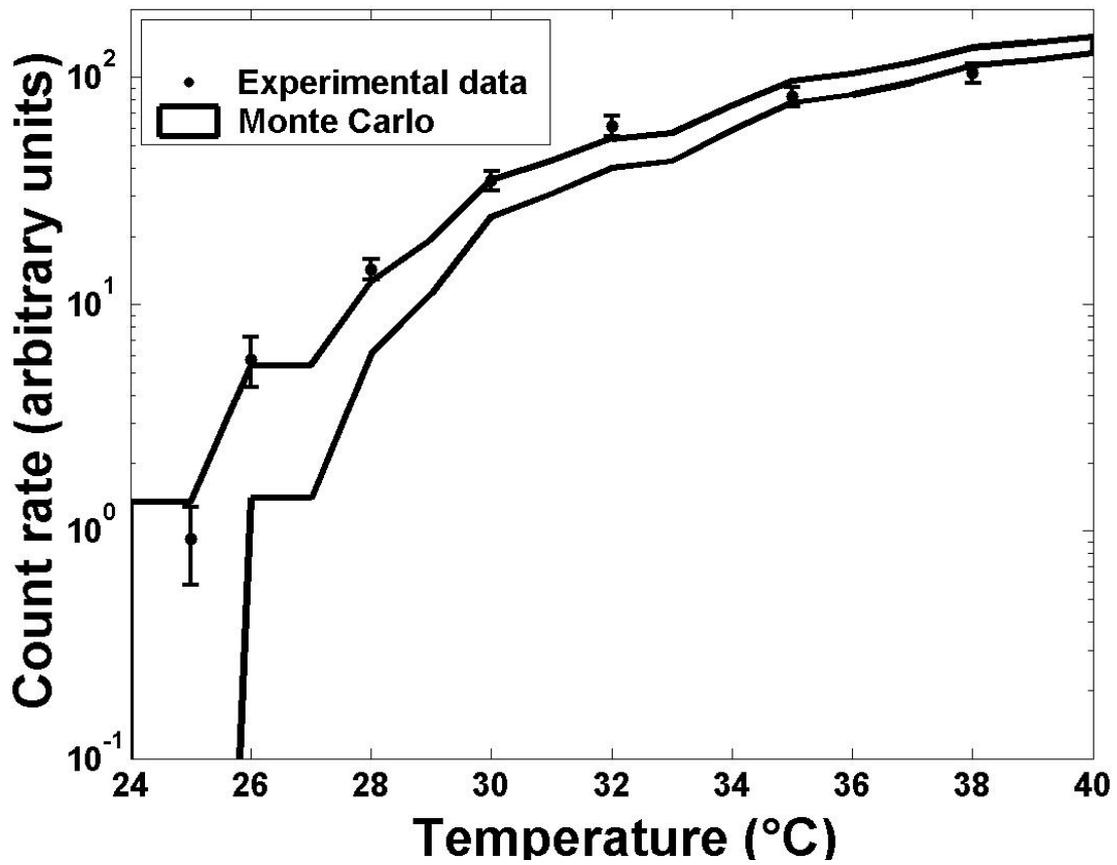

Figure 4: SBD-1000 response (count rate) to 400 keV neutrons as a function of temperature. The volume of the detector is 8 m$l$. The simulated response gives a loading of 0.68±0.02%.